\documentclass[12pt]{article}
\pdfoutput=1
\usepackage{amsmath,amssymb,cite}
\usepackage{graphicx}
\usepackage{color}
\usepackage[width=.8\textwidth,font=small,format=plain,labelfont=bf,up,textfont=it]{caption} 
\newcommand{\beq}{\begin{equation}}
\newcommand{\eeq}{\end{equation}}

\numberwithin{equation}{section}

\begin{document}

\vspace*{0.5 cm}

\begin{center}

{\Large{\bf NLO Leptoquark Production and Decay: }}

{\Large{\bf The Narrow-Width Approximation and Beyond. }}

\vspace*{1 cm}

{\large J.B. Hammett~$^1$, and D.A. Ross~$^1$} \\ [0.5cm]
{\it $^1$ School of Physics and Astronomy, University of Southampton,\\Highfield, Southampton SO17 1BJ, UK}\\[0.1cm]
 \end{center}

\vspace*{3 cm}

\begin{center}
{\bf Abstract} \end{center}
We study the  leptoquark model of Buchm\"uller, R\"uckl and 
Wyler, focusing on a particular type of scalar ($R_2$) and vector ($U_1$) leptoquark. The primary aim is to perform the 
calculations for leptoquark production and decay at next-to-leading order (NLO) to establish the importance of the NLO contributions 
and, in particular,  to determine how effective the narrow-width-approximation (NWA) is at NLO.
For both the scalar and vector leptoquarks it is found that the NLO contributions are large, with the larger corrections occurring
for the case vector leptoquarks. For the scalar leptoquark it is found that the NWA provides a good approximation for
determining the resonant peak, however the NWA is not as effective for the vector leptoquark.
For both the scalar and vector leptoquarks there are large contributions away from the resonant peak,
which are missing from the NWA results, and these make a significant difference to the total cross-section.


\vspace*{3 cm}

\begin{flushleft}
  Jun 2015 \\
\end{flushleft}

 \newpage
 
\section{Introduction}
For the purpose of this paper will use the leptoquark model of W.~Buchm\"uller, R.~R\"uckl and D.~Wyler \cite{BRW} - a model which conserves lepton and
baryon number thereby avoiding the problem of proton decay. The search for leptoquarks is an important consideration for the 
ATLAS \cite{ATLAS1,ATLAS2,ATLAS3} and CMS experiments \cite{CMS1,CMS2,CMS3} and the lepton and baryon number conservation feature of the 
Buchm\"uller, R\"uckl and Wyler model allows us to consider a leptoquark mass which is accessible to these LHC experiments.

A leptoquark production process can produce either a single leptoquark or a pair of leptoquarks, and in the study by Belyaev et. al. 
\cite{Belyaev} it has been shown that pair production dominates for low leptoquark masses ($m_\text{LQ}\sim100\,\text{GeV}$), but as the mass is 
increased single leptoquark production becomes the more dominant contribution (see also \cite{Dorsner} and \cite{Mandal}). 
For this reason we are focusing on single leptoquark production and 
the primary objective of this paper is to study single leptoquark production and decay at next-to-leading (NLO), in particular to determine the 
effectiveness of the narrow-width approximation (NWA) at NLO.
\footnote{The cross-section for leptoquark pair production at the LHC and
Tevatron have been studied by Kramer et. al. \cite{kramer}}
 In the case of scalar leptoquarks (within the NWA) it was found that electroweak 
corrections are negligible compared to the QCD corrections and therefore for the purpose of this study only QCD corrections will be considered for the 
NLO calculations.

The key assumptions of the NWA are that the leptoquark process can be factorised into a production and decay process (where the decay-width is much
smaller than the leptoquark mass) and that the interference between the different helicity states for the production and decay processes is negligible.
With these assumptions\footnote{Note: The second assumption only applies to vector leptoquarks, which have different helicity states.} the leptoquark 
production and decay can be calculated as separate processes involving an on-shell leptoquark. 

In addition to these assumptions, the studies by Kauer et. al. \cite{Kauer1,Kauer2,Kauer3} identify 
other factors which need to be considered when using the NWA. These include the effect of non-factorisable contributions and the break-down of the NWA 
when the decay involves a decay product with a mass that approaches that of the parent particle. The effect of non-factorisable contributions will be 
relevant to this paper, however the second consideration does not affect leptoquark production in the NWA since the leptoquark decay products are assumed to be 
massless in the high energy limit.

The NLO calculations are performed numerically using FORTRAN with the virtual corrections calculated using SAMURAI \cite{SAMURAI} and the cancellation 
of the infrared divergences implemented using the dipole subtraction method \cite{CataniSeymour,CDST}. In calculating the virtual corrections it was
found that some of the loop contributions introduced numerical instabilities requiring the calculations to be performed using quadruple precision within
FORTRAN.

The Buchm\"uller, R\"uckl and Wyler model includes both scalar and vector leptoquarks and these fall into two categories: those with 
fermion number
\footnote{Fermion number $F$ is defined as $F=3B+L$ where $B$ and $L$ are the baryon and lepton numbers respectively} $|F|=0$ and those with fermion 
number $|F|=2$. In this model the effective Lagrangian for the interactions of leptoquarks with leptons and quarks is
\beq \mathcal{L}_\text{quark+lepton} = \mathcal{L}_{|F|=2}+\mathcal{L}_{|F|=0} \eeq
with
\begin{align}
 \mathcal{L}_{|F|=2} &= (g_{1L}\,\overline{q}^c_L\, i \tau_2\, l_L + g_{1R}\,\overline{u}^c_R\, e_R)\,S_1 
                     + \tilde g_{1R}\,\overline{d}^c_R\,e_R\,\tilde S_1
                     + g_{3L}\,\overline{q}^c_L\,i\tau_2\mathbf{\tau}\,l_L\,S_3 \nonumber \\
                     &+ (g_{2L}\,\overline{d}^c_R\gamma^\mu\,l_L + g_{2R}\,\overline{q}^c_L\gamma^\mu\,e_R)\,V_{2\mu}
                     +\tilde g_{2L}\,\overline{u}^c_R\gamma^\mu\,l_L\,\tilde V_{2\mu} + \text{c.c.}
\end{align}
and
\begin{align}
 \mathcal{L}_{|F|=0} &= (h_{2L}\,\overline{u}_R\,l_L + h_{2R}\,\overline{q}_L\,i\tau_2\,e_R)\,R_2 + \tilde h_{2L}\,\overline{d}_R\,l_L\,\tilde R_2 \nonumber \\
                     &+ (h_{1L}\,\overline{q}_L\,\gamma^\mu\,l_L + h_{1R}\,\overline{d}_R\,\gamma^\mu\,e_R)\,U_{1\mu}
                     + \tilde h_{1R}\,\overline{u}_R\,\gamma^\mu e_R\,\tilde U_{1\mu} \nonumber \\
                     &+ h_{3L}\,\overline{q}_L\,\mathbf{\tau}\gamma^\mu\,l_L\, U_{3\mu} + \text{c.c.} \label{brwf=0}
\end{align}
where $q_L$ and $l_L$ are the left-handed quark and lepton doublets respectively and $e_R$, $u_R$, $d_R$ are the right handed charged leptons, u and
d-quarks respectively.

In this paper we will start by considering the $R_2$ scalar leptoquark, which has fermion number $|F|=0$ and is an isospin doublet with 
electric charges $+5/3$ and $+2/3$ and couples to the quark and lepton sector with couplings $h_{2L}$ and $h_{2R}$ - as detailed in eq.(\ref{brwf=0}). 
The leptoquark production and decay process will first be considered within the NWA and the results compared 
against the full non-factorisable process to determine the effectiveness of the NWA. This procedure will then be repeated with the $U_1$ vector 
leptoquark. The $U_1$ leptoquark also has fermion number $|F|=0$ and is an isospin singlet with electric charge $+2/3$ and couples to
the quarks and leptons with couplings $h_{1L}$ and $h_{1R}$ - also shown in eq.(\ref{brwf=0}).
Having considered both scalar and vector leptoquarks the paper will conclude with a discussion and comparison between the scalar and vector leptoquark 
results.

In addition to studying the effectiveness of the NWA at NLO this paper will also look at the sensitivity of the NLO results to the renormalisation and
factorisation scales.

In performing these calculations the center-of-mass energy has been set to $\sqrt{s}=14\,\text{TeV}$. The leptoquark mass has been chosen to be
$m_\text{LQ}=750\,\text{GeV}$ and the couplings have been set to
$\underbrace{h_{1L}=h_{1R}}_\text{$U_1$ couplings}=\underbrace{h_{2L}=h_{2R}}_\text{$R_2$ couplings}=1.07$ (i.e. taken from $\alpha_s$ evaluated
at $m_\text{LQ}$). The renormalisation and factorisation scales $\mu$ and $\mu_F$ have been set to $\mu=\mu_F=m_\text{LQ}$.

In section 2 we consider the case of scalar leptoquarks and vector leptoquarks in section 3. In section 4 we present our conclusions. 

\section{Scalar Leptoquarks}\label{scalarintro}
\begin{figure}[!ht]
 \begin{center}
  \includegraphics[scale=0.4]{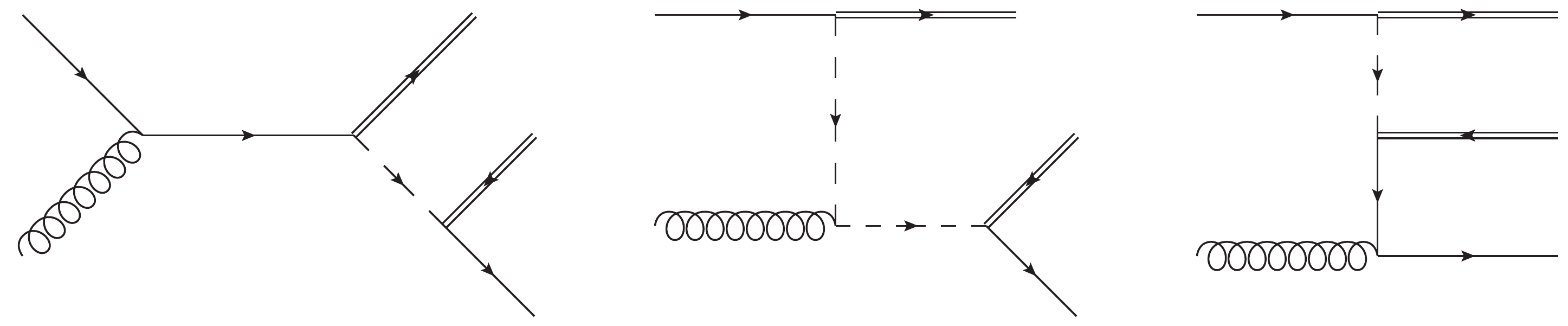}
  \caption{The LO contributions to the core process for leptoquark production. The first two diagrams can be factorised
into leptoquark production and decay amplitudes whereas the third diagram is non-factorisable.} 
  \label{fig:single_production}
 \end{center}
\end{figure}

For the study of scalar leptoquarks we will consider the $R_2$ type leptoquark as described in \cite{BRW}. The 
core process being studied is
 \beq u+g\rightarrow e^- + e^+ + u \label{core1} \eeq 
and the LO Feynman diagrams contributing to this process
are shown in figure \ref{fig:single_production}.

\begin{figure}[!ht]
 \begin{center}
  \includegraphics[scale=0.4]{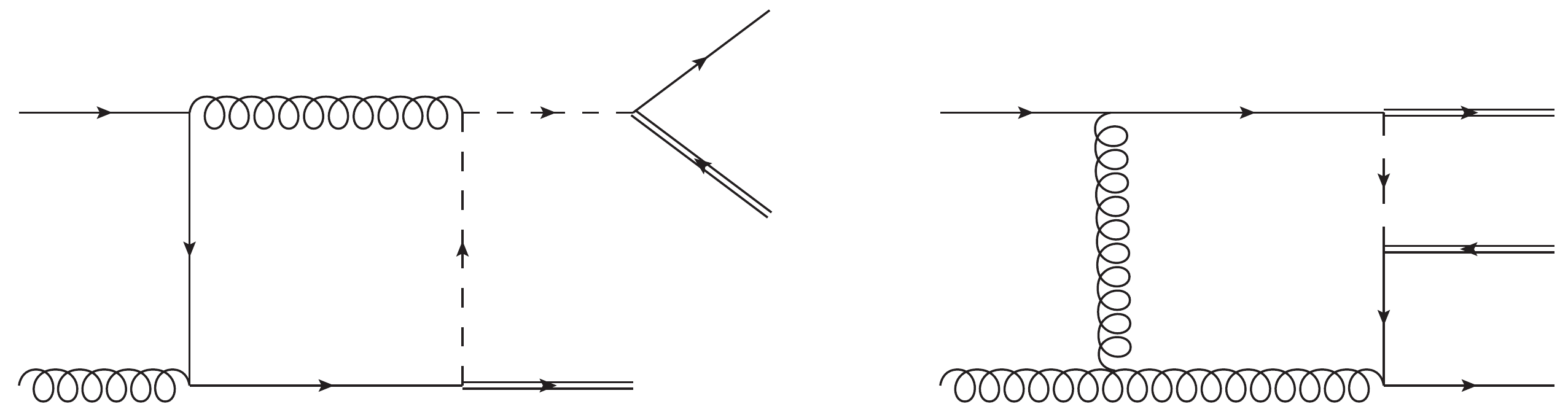}
  \caption{Example virtual corrections to the leptoquark production and decay process. The box diagram on the left is an example of
a factorisable process whereas the pentagon diagram on the right is non-factorisable.} 
  \label{fig:example_virtuals}
 \end{center}
\end{figure}

\begin{figure}[!ht]
 \begin{center}
  \includegraphics[scale=0.4]{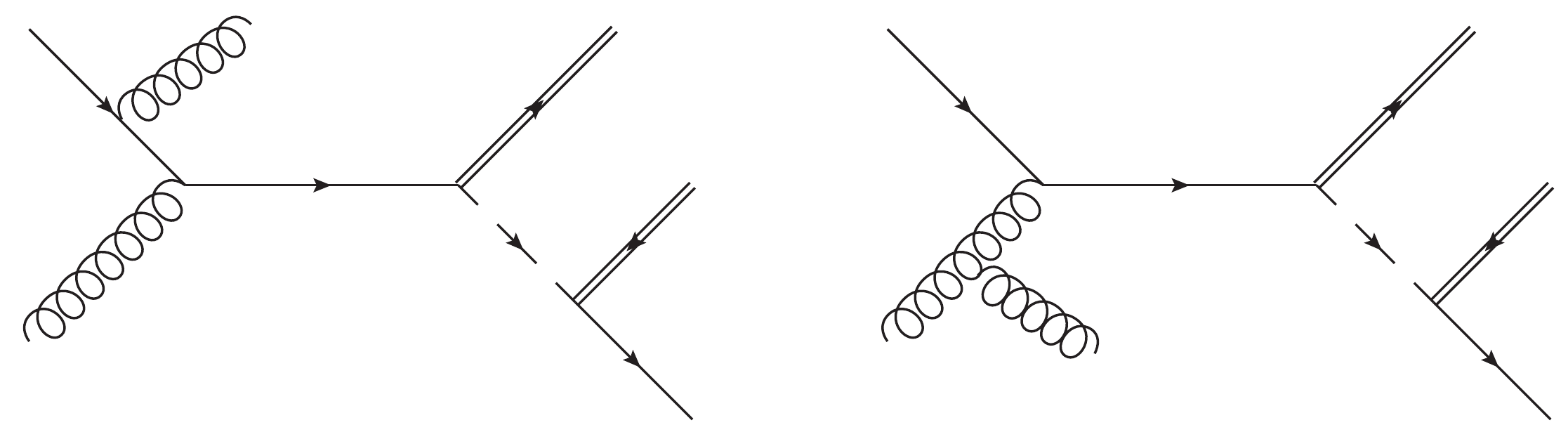}
  \caption{Example real corrections to the leptoquark production and decay process. The interference between these two diagrams 
contributes to the cancellation of the infrared divergences arising from the virtual corrections.} 
  \label{fig:example_reals}
 \end{center}
\end{figure}

\begin{figure}[!ht]
 \begin{center}
  \includegraphics[scale=0.4]{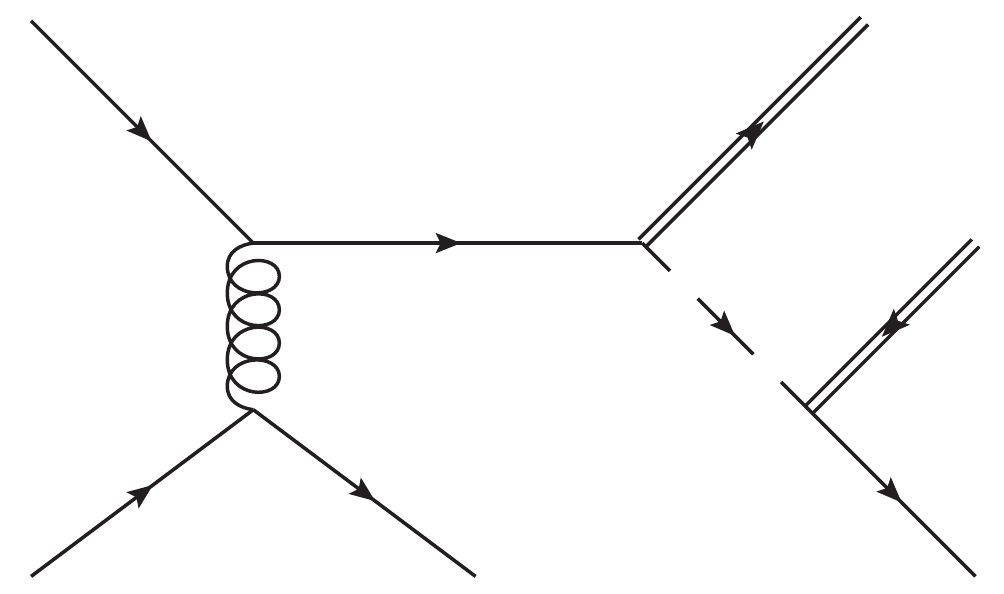}
  \caption{Example of an additional NLO contribution. This process shows leptoquark production from a quark-quark initial state.} 
  \label{fig:example_additional}
 \end{center}
\end{figure}

When studying leptoquark production at NLO there are three main contributions to the final result. Firstly there are the LO order 
contributions which come from the diagrams shown figure \ref{fig:single_production}, then there are the NLO virtual and real QCD
corrections to this process - with example diagrams shown in figures \ref{fig:example_virtuals} and \ref{fig:example_reals}. 
Finally at NLO there are additional initial state contributions to the leptoquark production process such as the quark-quark
initial state shown in figure \ref{fig:example_additional}.
There are similar contributions from quark-antiquark, antiquark-antiquark, and gluon-gluon scattering.
At the order of perturbation theory of the NLO corrections to the core process (\ref{core1}) these are given by 
tree-level amplitudes with an additional final-state parton.
 Some of the additional NLO calculations suffer from initial state singularities. There is no corresponding
 virtual correction with a cancelling infrared divergence - the divergences are absorbed into the parton distribution functions
 (PDF) and in keeping with our treatment of the infrared singularities in the NLO corrections to the core process, 
 they are handled using the dipole subtraction method \cite{CataniSeymour,CDST}.

In addition to the above mentioned contributions there are also background diagrams to consider.
The main background to this process arises from Drell-Yan pair production in the presence of an accompanying jet. The cross-section for this process 
is considerably smaller than the total Drell-Yan cross-section.  Furthermore, away from the resonant region for the leptoquark we expect the 
relative contribution from this process to be negligible as it does not contribute to a resonance in the invariant mass of the lepton-jet system. 
Since this process involves the emission of an intermediate $Z$ or photon, we expect that even away from the resonance region, the cross-section will 
be suppressed by the fourth power ratio of the electroweak coupling to the coupling of the leptoquark, which we take to be of the order of the
strong coupling.  The amplitude for the Drell-Yan process rapidly decreases with increasing invariant mass of the lepton pair (above the $Z$-mass 
resonance) and so any remaining background interference can be eliminated by imposing an appropriate minimum cut on the lepton-pair invariant mass.
For this reason the contributions from the additional background diagrams have been excluded from this study.

All of the results in this paper show the differential cross-section versus the invariant mass, $m_\text{inv}$, of the jet + anti-lepton system
evaluated over the range $[0,2m_\text{LQ}]$. In the case of one jet the invariant mass is defined as $m^2_\text{inv} \equiv
\left(p_{e^+}+p_{\text{jet}}\right)^2$
where $p_{e^+}$ is the 4-momentum of the anti-lepton and $p_\text{jet}$ is the 4-momentum of the jet. When there are two jets in the final state
 there are two such invariant masses,
$m_{\text{inv}^{(1)}}$ and $m_{\text{inv}^{(2)}}$ where 
 $m^2_{\text{inv}^{(i)}} \equiv
\left(p_{e^+}+p_{\text{jet}_i}\right)^2$, \ $i=1,2$. 
When constructing the differential cross-section w.r.t. $m_\text{inv}$
we include in a given bin of $m_\text{inv}$ any event in which {\it either}
of the invariant masses lies within that bin. 
As it is not  possible to 
distinguish experimentally between the two jets, in order to determine
 which jet arises from the decay of a leptoquark, an event is deemed to lie
 within a certain bin if it contains a positron and any jet  with
invariant mass within that bin.
In the resonant region, the contribution from the "wrong" jet (i.e. that which did not arise from the leptoquark decay) 
will make a very small relative contribution to the differential cross-section.

\subsection{NWA Results}
\begin{figure}[!ht]
 \begin{center}
  \includegraphics[scale=0.55]{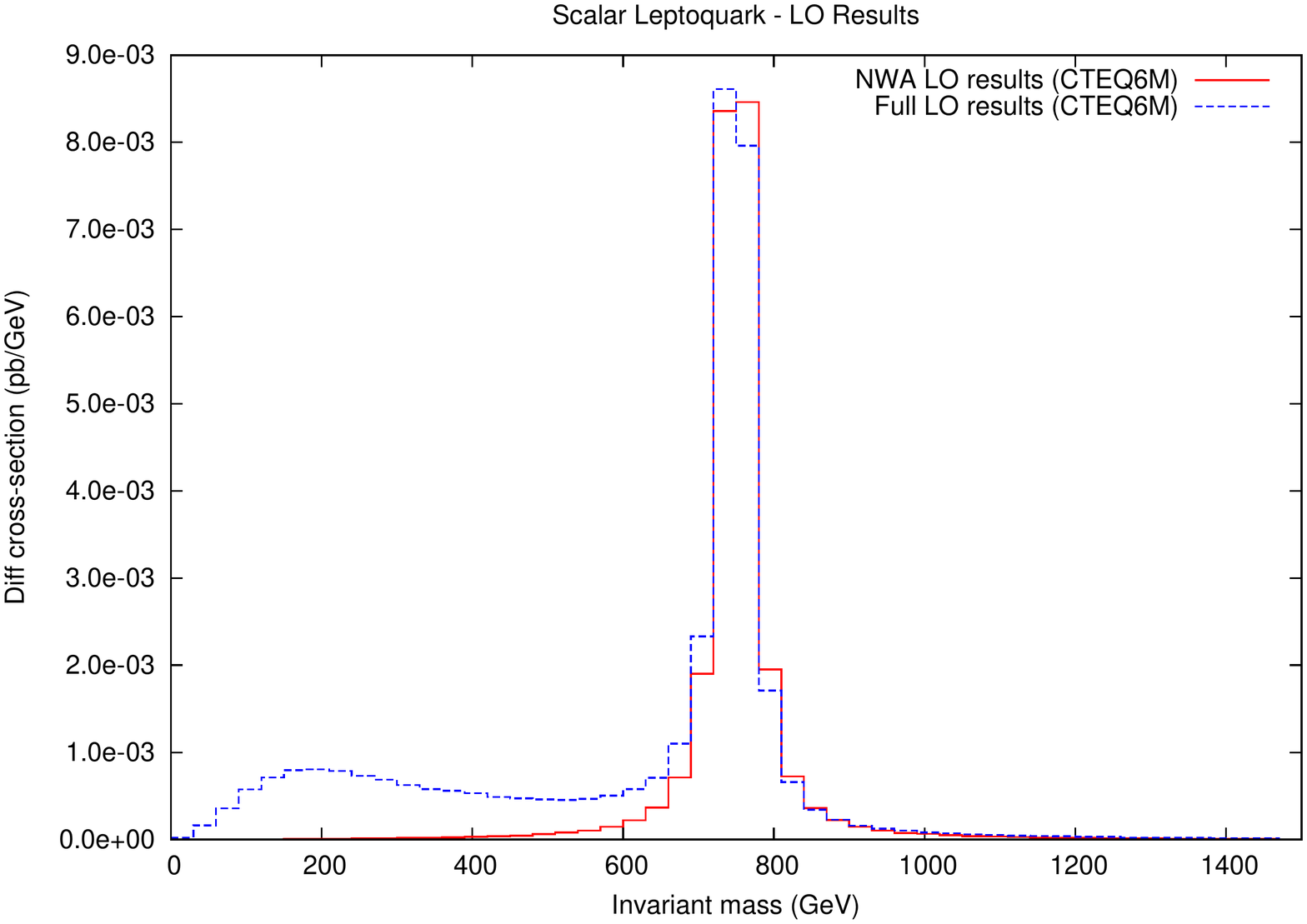}
  \caption{LO results for producing a scalar leptoquark - comparing the NWA to the full non-factorisable process (CTEQ6M).
The dashed line includes the effect of the non-factorising  graph at LO. } 
  \label{fig:scalar_tree_graph}
 \end{center}
\end{figure}

\begin{figure}[!ht]
 \begin{center}
  \includegraphics[scale=0.55]{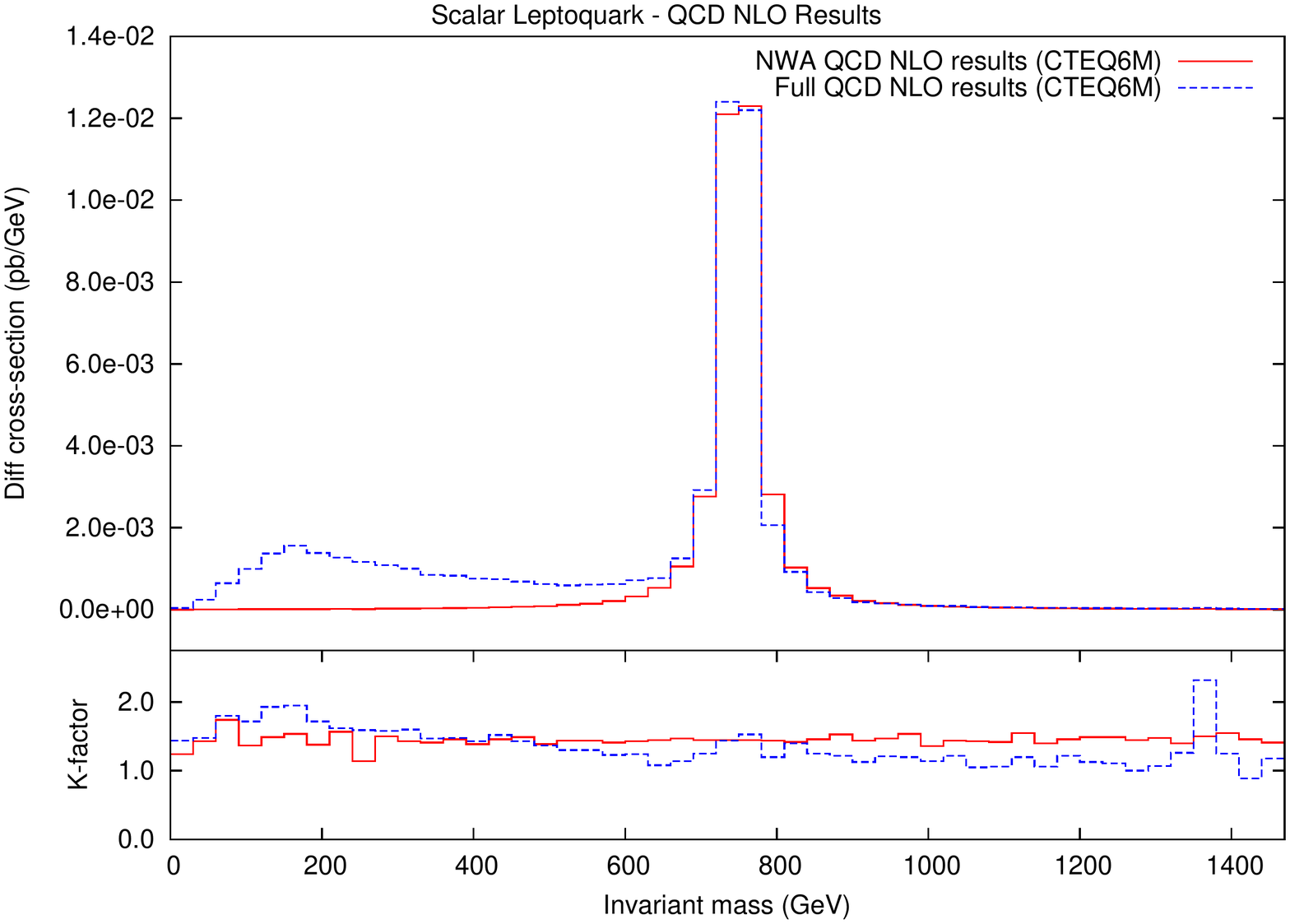}
  \caption{NLO results for producing a scalar leptoquark from the core partonic process - comparing the NWA to the full non-factorisable process (CTEQ6M). These results
include the QCD corrections only.} 
  \label{fig:scalar_qcd_graph}
 \end{center}
\end{figure}

\begin{figure}[!ht]
 \begin{center}
  \includegraphics[scale=0.55]{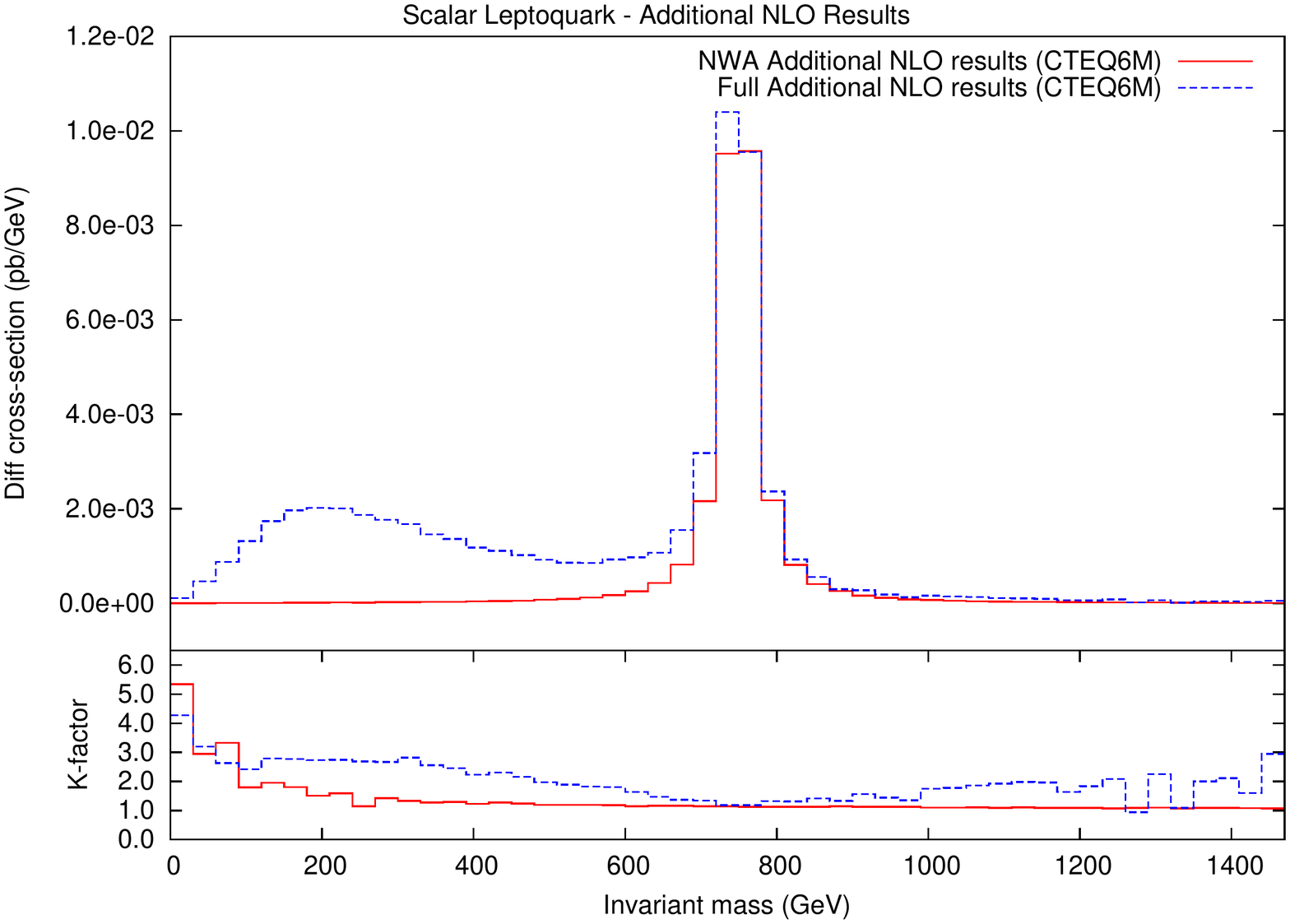}
  \caption{Additional NLO results (i.e. from the additional partonic processes) for producing a scalar leptoquark - comparing the NWA to the 
  full non-factorisable process (CTEQ6M). These results include QCD corrections only.} 
  \label{fig:scalar_add_graph}
 \end{center}
\end{figure}

\begin{figure}[!ht]
 \begin{center}
  \includegraphics[scale=0.55]{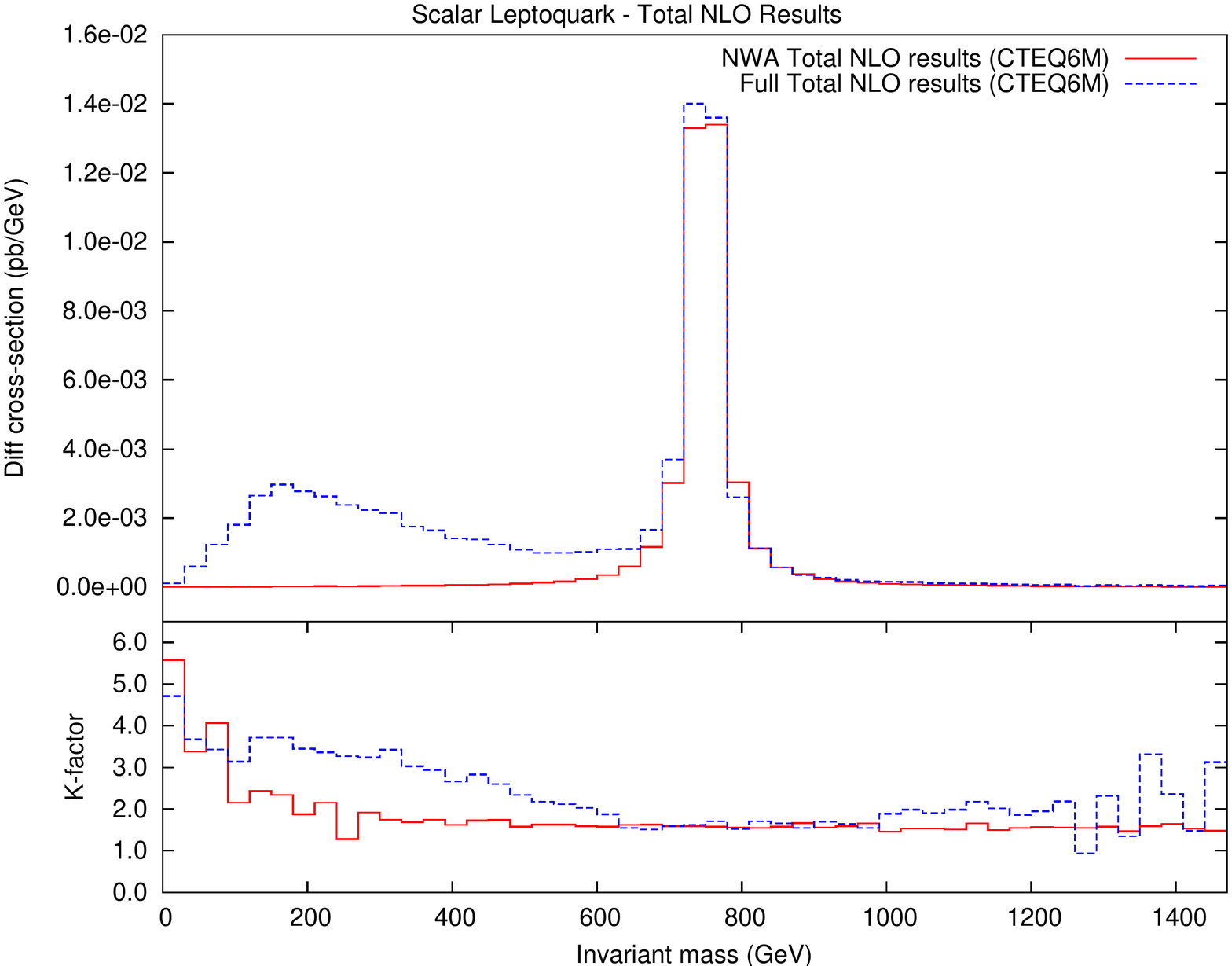}
  \caption{Total NLO results for producing a scalar leptoquark - comparing the NWA to the full non-factorisable process (CTEQ6M).} 
  \label{fig:scalar_total_graph}
 \end{center}
\end{figure}

\begin{figure}[!ht]
 \begin{center}
  \includegraphics[scale=0.45]{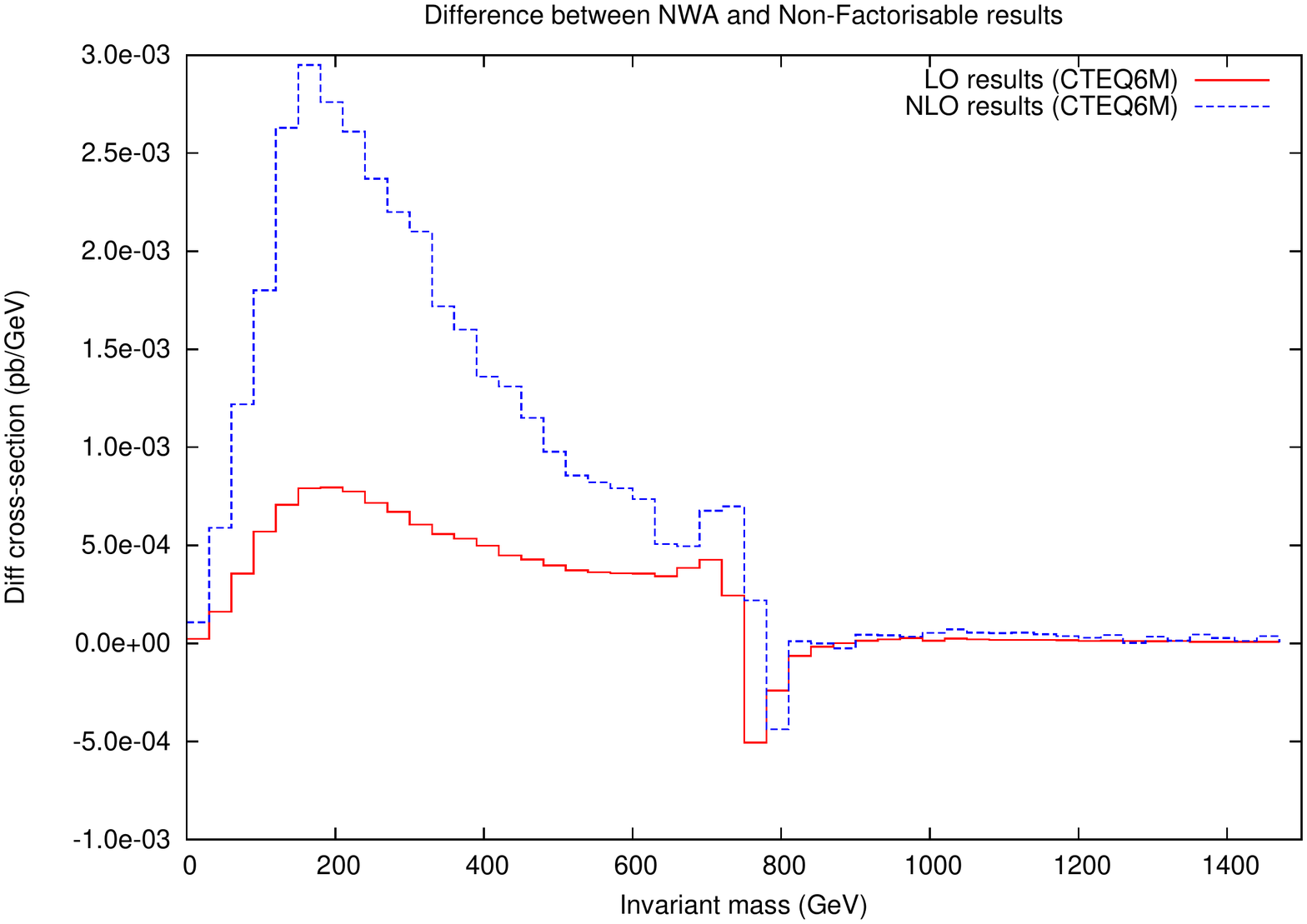}
  \caption{Difference between the NWA and full non-factorisable scalar leptoquark results at LO and NLO.} 
  \label{fig:scalar_diff_graph}
 \end{center}
\end{figure}

Looking first at the NWA results, figure \ref{fig:scalar_tree_graph} shows the LO contribution. As can be seen from the plot the 
NWA provides a symmetrical resonance with a peak near $m_\text{LQ} = 750\,\text{GeV}$ with a height of
$8.46\times10^{-3}\,\text{pb/GeV}$. The total cross-section at LO is $0.74\,\text{pb}$·

Referring to figure \ref{fig:scalar_qcd_graph} the NLO QCD corrections to the core 
parton process,(\ref{core1}), make a significant contribution to the process. There is still a 
symmetric resonant peak, but now with a height of $1.23\times10^{-2}\,\text{pb/GeV}$ which gives an enhancement of $45\%$ over 
the LO result. For the QCD corrections the total cross-section is $1.07\,\text{pb}$ giving an increase of $45\%$ over the LO result.
The K-factor for the NLO QCD corrections is reasonably constant across the invariant mass distribution with an average value
$\sim 1.5$

Finally the  contributions from the additional partonic processes are shown in figure \ref{fig:scalar_add_graph}.
As discussed in section \ref{scalarintro} the additional partonic processes are the NLO tree-level amplitudes which involve
different initial state partons such as the quark-quark initial state shown in figure \ref{fig:example_additional}.
These additional contributions also enhance the LO process. The height of
the resonant peak is $9.57\times10^{-3}\,\text{pb/GeV}$ giving an enhancement of $13\%$ over the LO result. The total cross-section
is $0.84\,\text{pb}$ which gives an increase over the LO result of $14\%$.
The K-factor for the additional partonic processes is $\sim 1.0$ across the invariant mass distribution, but does show an 
increased K-factor for low values of the invariant mass.

The combined results are shown in figure \ref{fig:scalar_total_graph} and give the total NLO contributions to the leptoquark production 
process in the NWA. The total NLO contributions give a resonant peak height of $1.34\times10^{-2}\,\text{pb/GeV}$ adding a sizeable 
enhancement of $58\%$ to the LO result. The total cross-section at NLO is $1.17\,\text{pb}$ which gives a $59\%$ increase over the 
LO result.\footnote{Similar large corrections to single leptoquark production  have been reported by Plehn et. al. \cite{plehn} and by Alves .et .al. 
\cite{alves}.}
The K-factor for the total NLO contributions is $\sim 1.5$ across the invariant mass distribution and shows an increased K-factor
for low values of the invariant mass coming from the additional partonic processes.
These results are summarised in tables \ref{table:scalar_nwa_peak} and \ref{table:scalar_nwa_cs}.

\begin{table}[!ht]
 \begin{center}
 \begin{tabular}{|l|r|r|}
  \hline
  Correction Type & Peak Height (pb/GeV) & Percentage change on LO \\ \hline
  LO & $8.46\times10^{-3}$ & - \\ \hline
  NLO QCD & $1.23\times10^{-2}$ & +45\% \\ \hline
  NLO Additional & $9.57\times10^{-3}$ & +13\% \\ \hline\hline
  NLO Total & $1.34\times10^{-2}$ & +58\% \\
  \hline
 \end{tabular}
 \caption{Summary of the CTEQ6M scalar results in the NWA.}
 \label{table:scalar_nwa_peak}
 \end{center}
\end{table}
\begin{table}[!ht]
 \begin{center}
 \begin{tabular}{|l|r|r|}
  \hline
  Correction Type & Total cross-section (pb) & Percentage change on LO \\ \hline
  LO & 0.74 & - \\ \hline
  NLO QCD & 1.07 & +45\% \\ \hline
  NLO Additional & 0.84 & +14\% \\ \hline\hline
  NLO Total & 1.17 & +59\% \\
  \hline
 \end{tabular}
 \caption{Summary of the CTEQ6M scalar results in the NWA.}
 \label{table:scalar_nwa_cs}
 \end{center}
\end{table}

\subsection{Non-Factorisable Results}
Figure \ref{fig:scalar_tree_graph} shows the contribution of the LO non-factorizing graph of figure \ref{fig:single_production}.
In common with the NWA result the main feature in the distribution is the resonance with a peak near $m_\text{LQ}=750\,\text{GeV}$. This has a peak 
height of $8.61\times10^{-3}\,\text{pb/GeV}$ which is very close to the NWA result. 
The key difference between the inclusion of the non-factorizing graph and the NWA result is that the distribution is no longer symmetric around the 
peak and we see that the non-factorisable contributions give an enhancement to the distribution for values of the invariant mass 
$m_\text{inv}<m_\text{LQ}$. The total cross-section at LO is $1.08\,\text{pb}$ in comparison to the NWA the total cross-section 
is larger. This is due to the enhancement to the distribution away from the resonance.

The results for the QCD corrections to the core process, (\ref{core1}), are shown in 
figure \ref{fig:scalar_qcd_graph} and give a similar distribution to the LO results, with an 
enhancement to the resonant peak. The peak height is increased by $44\%$ to $1.24\times10^{-2}\,\text{pb/GeV}$ and comparing this 
with the NWA peak the peaks are again very close. For the QCD corrections the total cross-section is $1.56\,\text{pb}$ which again 
is larger than the cross-section in the NWA and gives an enhancement over the LO cross-section of $45\%$.
The K-factor for the QCD contributions is reasonably constant over the invariant mass distribution, but shows more variation than
the NWA result.

The contributions at  NLO from the additional partonic processes  are shown in figure \ref{fig:scalar_add_graph}. This also gives a similar distribution to the LO result, but in 
addition to an enhancement to the resonant peak there is also a further enhancement in the region 
$m_\text{inv}<m_\text{LQ}$. The resonant peak has a height of $1.02\times10^{-2}\,\text{pb/GeV}$ giving an enhancement over
the LO result of $18\%$. Again comparing the peak height with the peak in the NWA the
peak heights are close. The total cross-section with the additional NLO results is $1.75\,\text{pb}$ and is 
considerably larger than the cross-section from NWA.
The K-factor for the additional partonic processes shows more variation than the NWA result. As with the NWA results there is an 
increase in the K-factor for both low and high values of the invariant mass distribution.

Combining the results gives the total NLO contribution to the leptoquark process and is shown in figure 
\ref{fig:scalar_total_graph}. The height of the resonant peak is $1.40\times10^{-2}$ which gives an enhancement of $62\%$
compared with the LO peak. There is also a further enhancement in the region $m_\text{inv}<m_\text{LQ}$ which is primarily due to
the  contributions from additional partonic processes at NLO. The cross-section for the total NLO result is $2.24\,\text{pb}$ which is
an increase of $107\%$ over the LO result, this is primarily due to the additional NLO contributions.
The K-factor for the total NLO contributions coincides with the K-factor for the NWA result around the resonant peak, but shows an
increase (with some fluctuation) at both low and high values of the invariant mass distribution.
A summary of these results is shown in tables \ref{table:scalar_full_peak} and  \ref{table:scalar_nwa_cs}.

\begin{figure}[!ht]
 \begin{center}
  \includegraphics[scale=0.75]{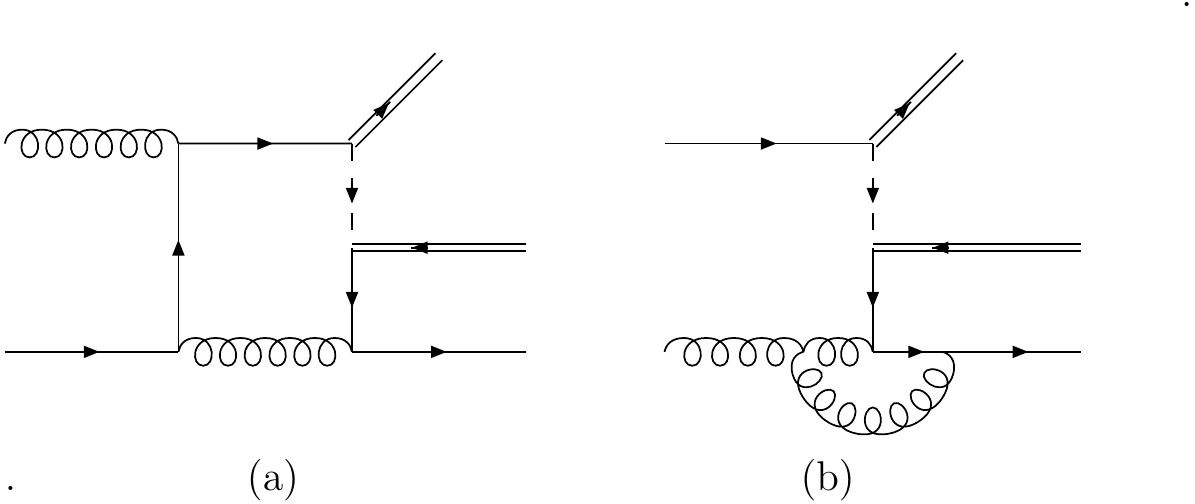}
  \caption{Graph (a) is an example of a non-factorizing virtual correction
 in which one end of the internal gluon  is attached to an incoming
 parton whereas
 the other end is attached to one of the leptoquark decay products.
NLO. Graph (b) is a virtual correction to the additional contribution present
 at tree-level  beyond the NWA in which a leptoquark is exchanged in  the 
 $t$-channel.} 
  \label{fig:nlo_graphs}
 \end{center}
\end{figure}

At NLO there is considerable enhancement, partly due to non-factorising 
one-loop corrections in  which a virtual gluon is exchanged between
 incoming  and outgoing partons (an example of which is shown in 
Fig.\ref{fig:nlo_graphs}(a)) and partly due to loop corrections
 to tree-level processes beyond the NWA  in which a leptoquark is 
exchanged in the $t$-channel (an example of which is shown in 
Fig.\ref{fig:nlo_graphs}(b)). This latter contribution gives rise to
 substantial enhancement below resonance. Note however that not all
 Feynman graphs can be associated unambiguously into one of the two 
above-mentioned corrections. For example, the graph shown on the RHS of 
Fig.\ref{fig:example_virtuals} can be interpreted as
 {\it either} an NLO correction to the third graph of
 Fig.\ref{fig:single_production} {\it or} a non-factorisable correction
 to the first graph of  Fig.\ref{fig:single_production}.
To highlight these contributions
figure \ref{fig:scalar_diff_graph} makes a comparison between the NWA and non-factorisable results for a scalar leptoquark by 
showing the difference between the two sets of results at both LO and NLO. For both the LO and NLO results there are small 
differences in the resonant region, but the biggest difference is seen in the region $m_\text{inv}<m_\text{LQ}$ coming from the 
enhancement to the distribution due to the non-factorisable process. The NLO results show the biggest difference in this region 
which, as discussed, is primarily due to the additional NLO contributions. 

\begin{table}[!ht]
 \begin{center}
 \begin{tabular}{|l|r|r|}
  \hline
  Correction Type & Peak Height (pb/GeV) & Percentage change on LO \\ \hline
  LO & $8.61\times10^{-3}$ & - \\ \hline
  NLO QCD & $1.24\times10^{-2}$ & +44\% \\ \hline
  NLO Additional & $1.02\times10^{-2}$ & +18\% \\ \hline\hline
  NLO Total & $1.40\times10^{-2}$ & +62\% \\
  \hline
 \end{tabular}
 \caption{Summary of the CTEQ6M scalar results for the non-factorisable process.}
 \label{table:scalar_full_peak}
 \end{center}
\end{table}
\begin{table}[!ht]
 \begin{center}
 \begin{tabular}{|l|r|r|}
  \hline
  Correction Type & Total cross-section (pb) & Percentage change on LO \\ \hline
  LO & 1.08 & - \\ \hline
  NLO QCD & 1.56 & +45\% \\ \hline
  NLO Additional & 1.75 & +62\% \\ \hline\hline
  NLO Total & 2.24 & +107\% \\
  \hline
 \end{tabular}
 \caption{Summary of the CTEQ6M scalar results for the non-factorisable process.}
 \label{table:scalar_full_cs}
 \end{center}
\end{table}

\subsection{Renormalization/Factorisation Scale Dependence for Scalar Leptoquarks}\label{section:scalar_scale_dependence}
Performing the calculations at NLO introduces renormalisation and factorisation scales - $\mu$ and $\mu_F$ respectively. In principal the choice of 
the values of these scales is arbitrary, however due to the nature of perturbative calculations the NLO corrections will have some sensitivity to 
$\mu$ and $\mu_F$. To determine the sensitivity of the results the NLO cross-section has been calculated for a range of different renormalisation 
and factorisation scales in the range $[\frac{1}{2}\,m_\text{LQ},2\,m_\text{LQ}]$ where we have limited ourselves
 to the case in which these two scales are equal.

\begin{figure}[!ht]
 \begin{center}
  \includegraphics[scale=0.55]{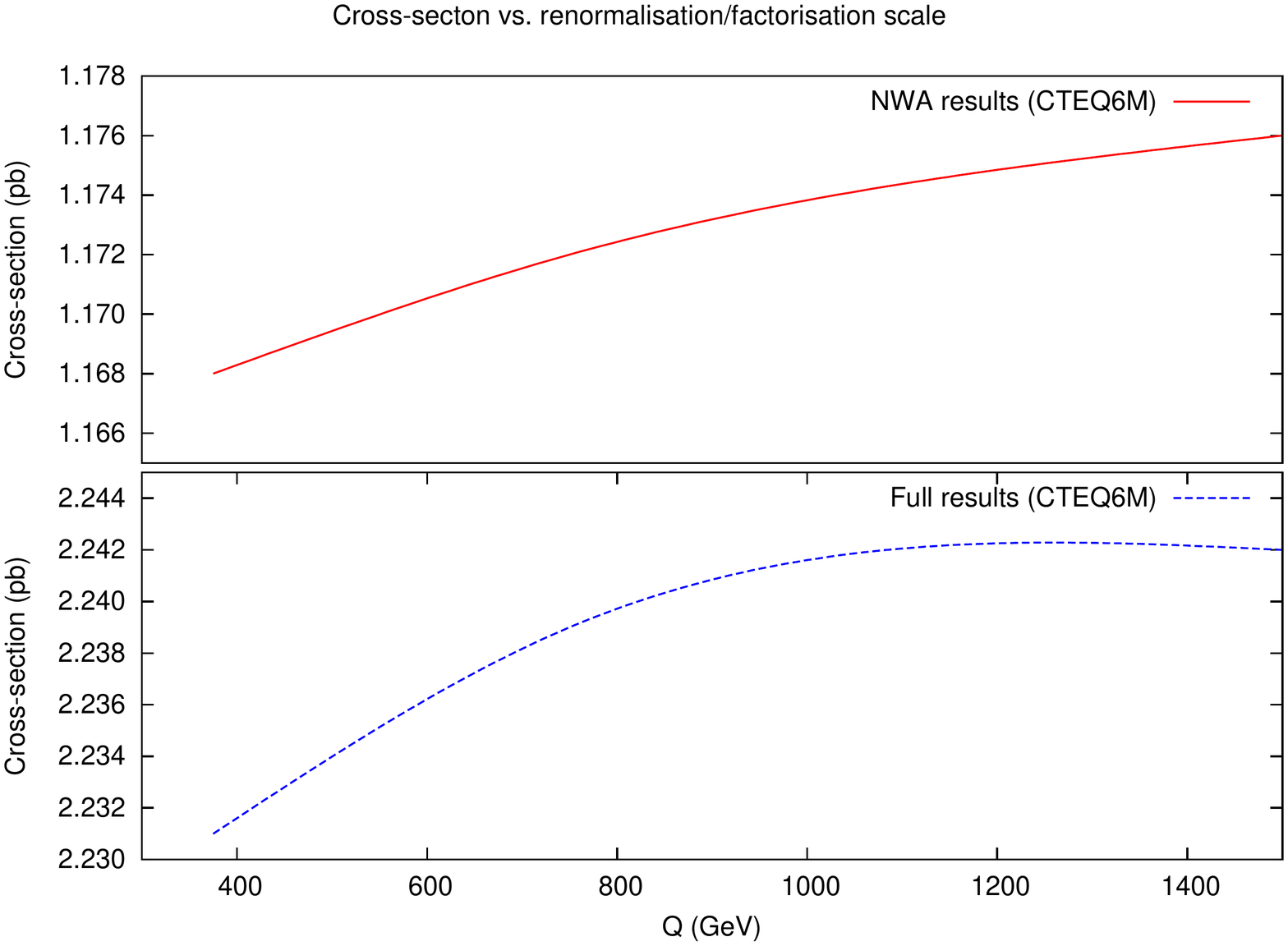}
  \caption{Scale dependence for the NWA and full non-factorisable scalar leptoquark process - with the scale $Q=\mu=\mu_F$.} 
  \label{fig:scalar_scale_graph}
 \end{center}
\end{figure}

The results of the scale dependence are given in figure \ref{fig:scalar_scale_graph} and show that, although
there is some scale dependence, the NLO results are reasonably insensitive to a change in $\mu$ and $\mu_F$.

\section{Vector Leptoquarks}
For the study of vector leptoquarks we will consider the $U_1$ type leptoquark described in \cite{BRW}. The core process
being studied is
 \beq d+g\rightarrow e^- + e^+ + d \label{core2} \eeq 
and the Feynman diagrams for this process at LO and NLO have the same topologies as in the scalar case.\footnote{The virtual corrections present one 
complication due to our choice of gauge for the vector particles: the Feynman gauge. In this gauge contributions from Goldstone bosons and 
Faddeev-Popov ghosts need to be included with the loop diagrams.}

As before the results are comprised of three main contributions. The LO contributions, the NLO virtual and real QCD corrections
to the LO process and the additional NLO tree-level contributions. Again the initial state singularities are handled using the dipole subtraction method.

\subsection{NWA Results}
\begin{figure}[!ht]
 \begin{center}
  \includegraphics[scale=0.55]{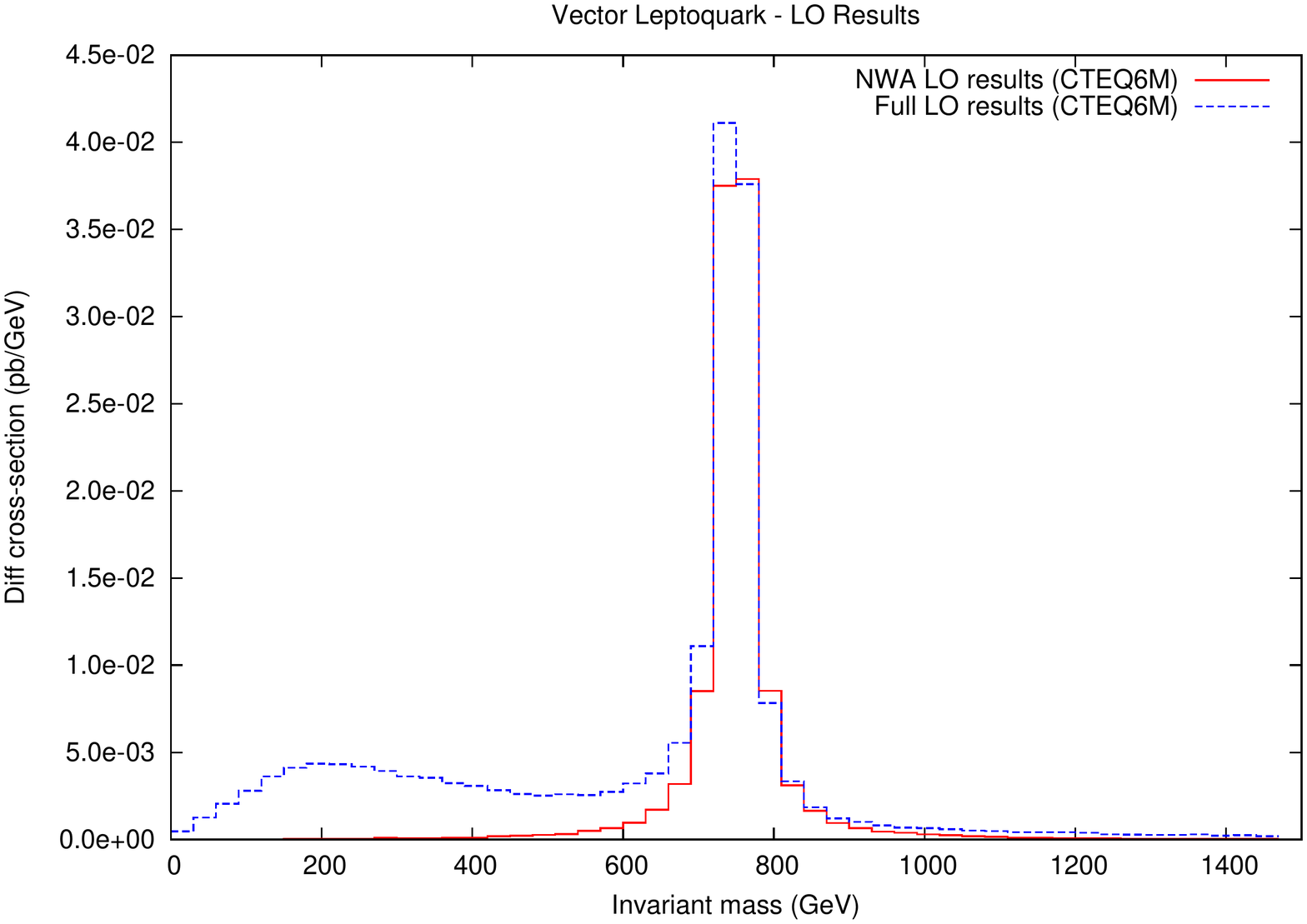}
  \caption{LO results for producing a vector leptoquark - comparing the NWA to the full non-factorisable process (CTEQ6M).} 
  \label{fig:vector_tree_graph}
 \end{center}
\end{figure}

\begin{figure}[!ht]
 \begin{center}
  \includegraphics[scale=0.55]{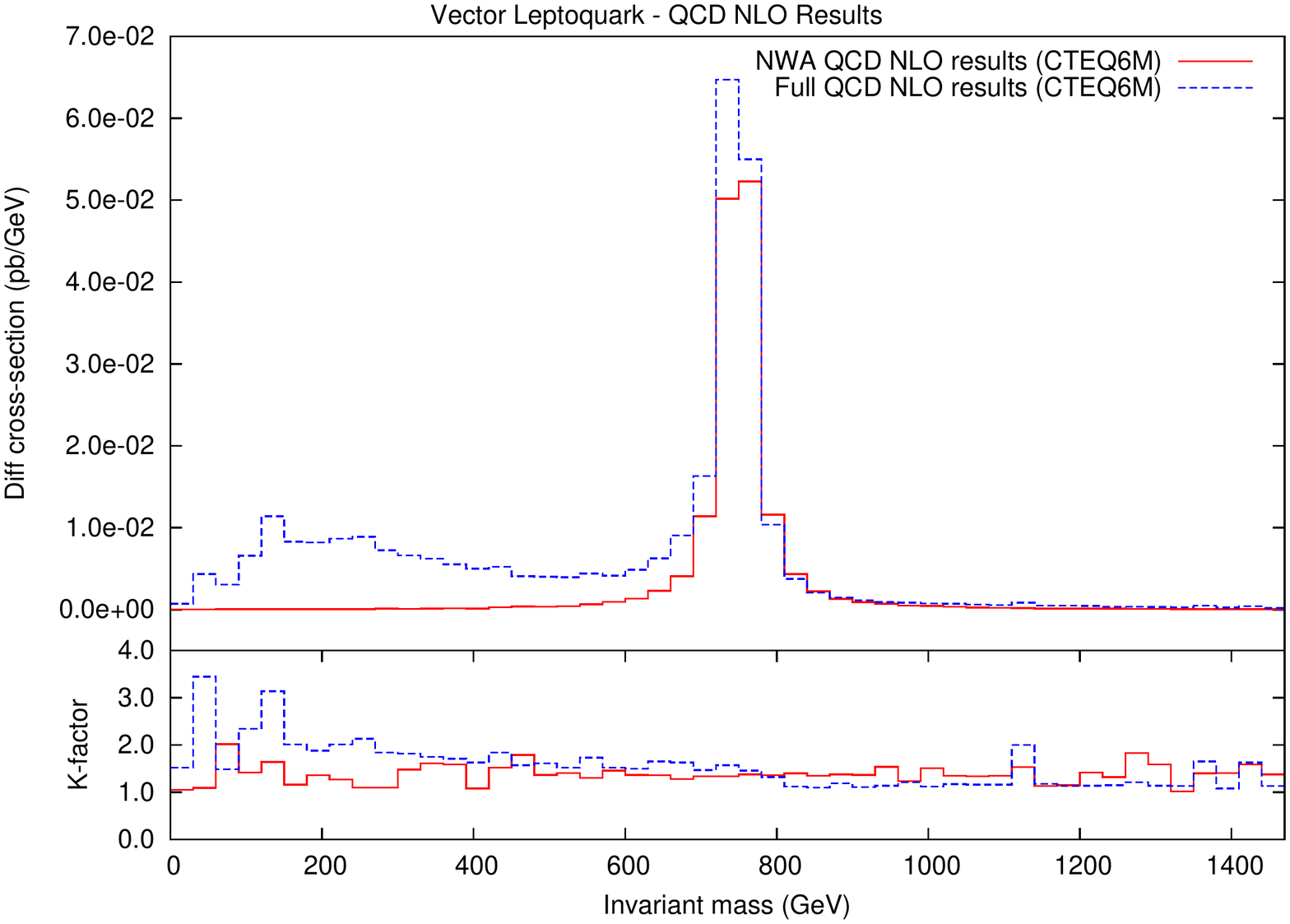}
  \caption{NLO results for producing a vector leptoquark - comparing the NWA to the full non-factorisable core process (CTEQ6M). These results
include the QCD corrections only.} 
  \label{fig:vector_qcd_graph}
 \end{center}
\end{figure}

\begin{figure}[!ht]
 \begin{center}
  \includegraphics[scale=0.55]{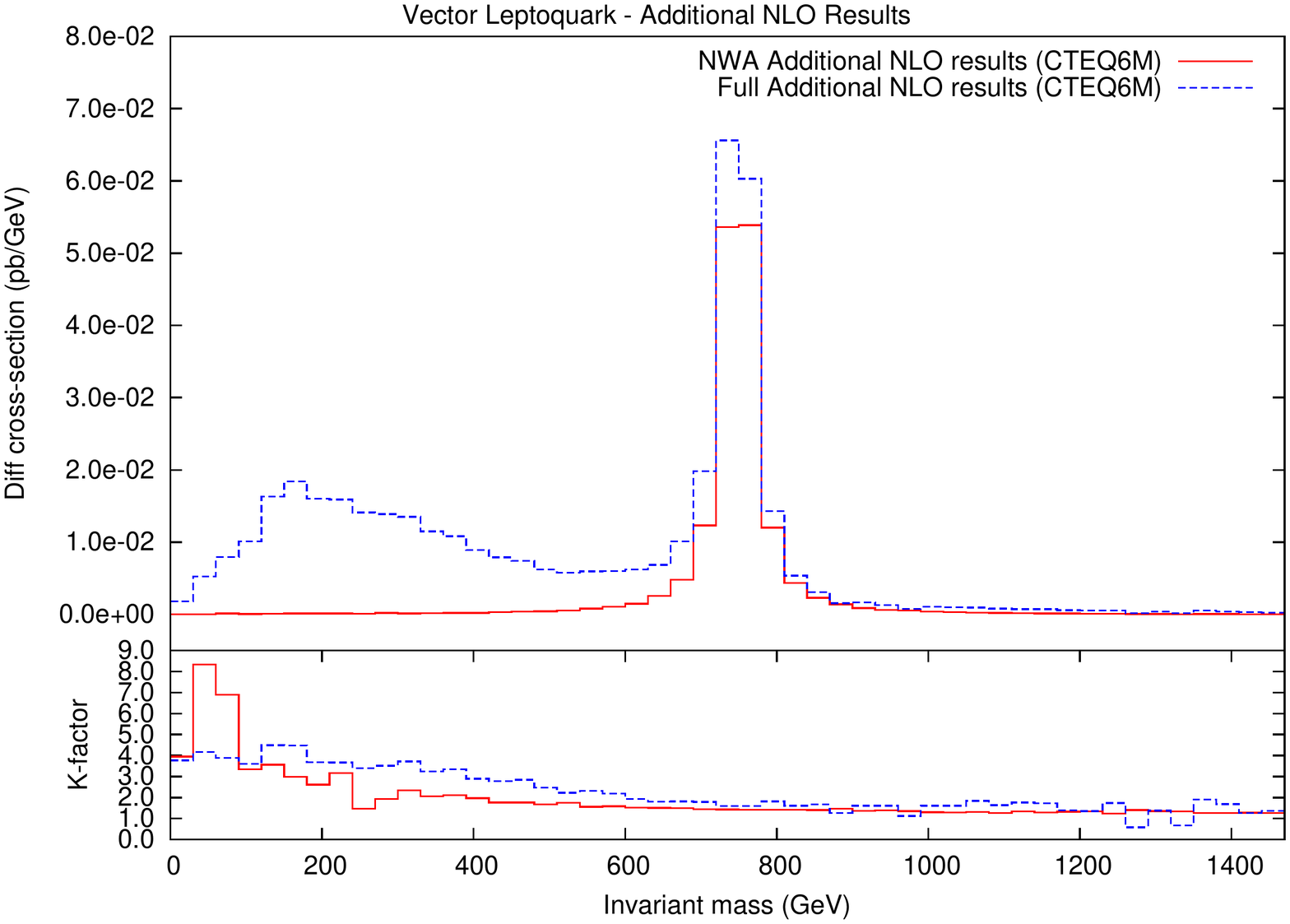}
  \caption{Contributions from the additional partonic processes at NLO for producing a vector leptoquark - comparing the NWA to the full non-factorisable process (CTEQ6M). These 
results include QCD corrections only.} 
  \label{fig:vector_add_graph}
 \end{center}
\end{figure}

\begin{figure}[!ht]
 \begin{center}
  \includegraphics[scale=0.55]{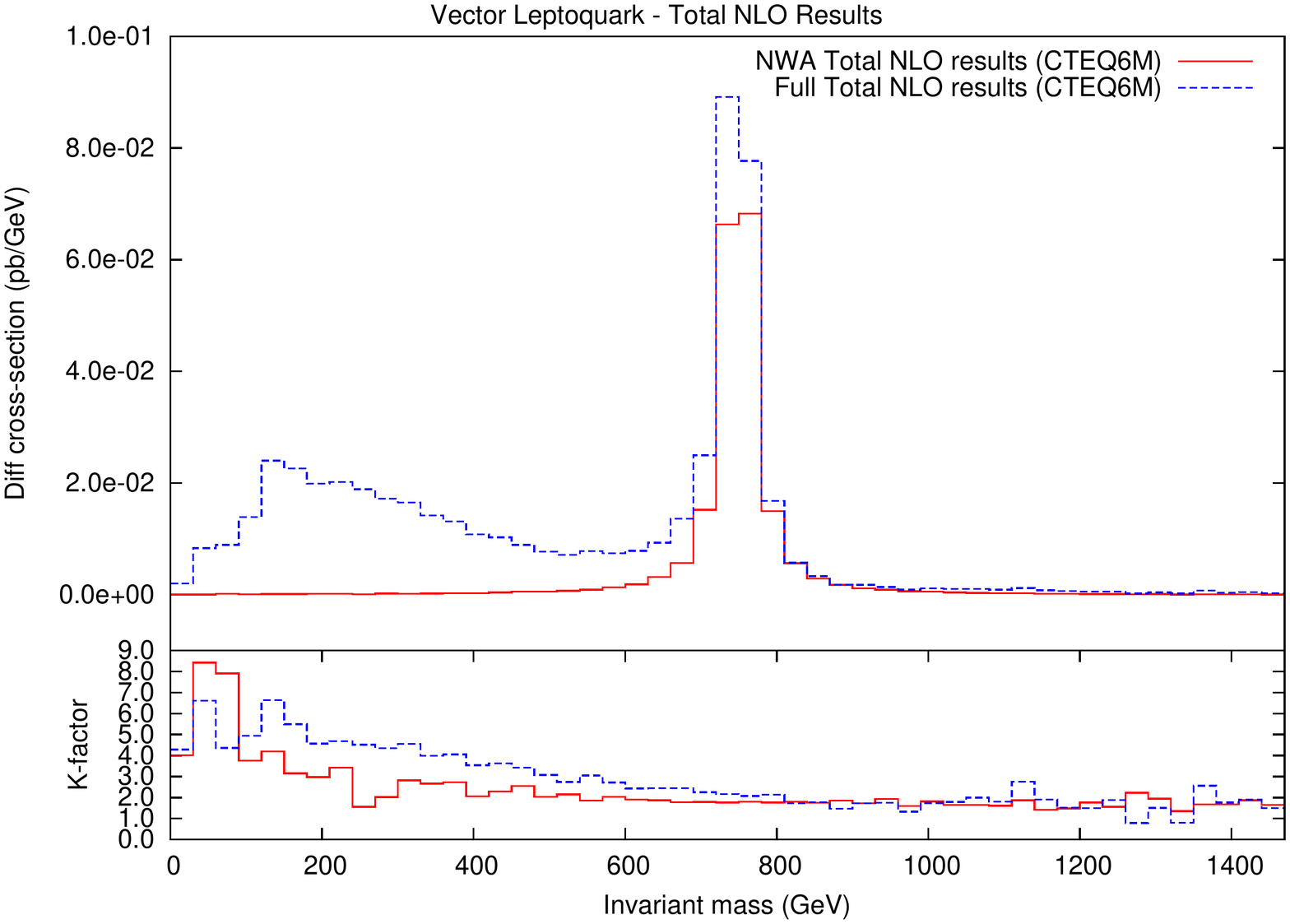}
  \caption{Total NLO results for producing a vector leptoquark - comparing the NWA to the full non-factorisable process (CTEQ6M).} 
  \label{fig:vector_total_graph}
 \end{center}
\end{figure}

\begin{figure}[!ht]
 \begin{center}
  \includegraphics[scale=0.45]{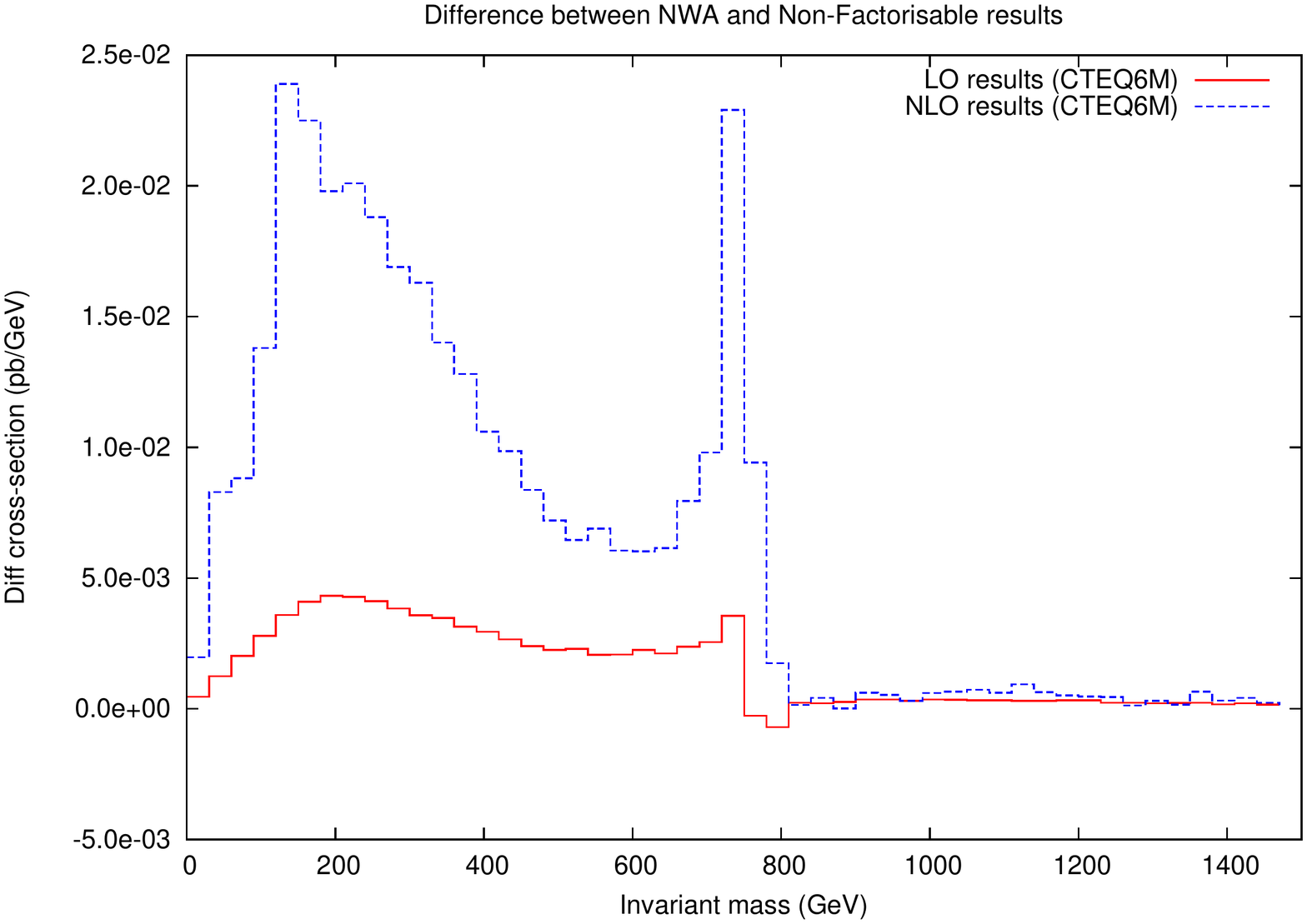}
  \caption{Difference between the NWA and full non-factorisable vector leptoquark results at LO and NLO.} 
  \label{fig:vector_diff_graph}
 \end{center}
\end{figure}

The NWA result for the vector leptoquark show the same features as were seen with the scalar leptoquark. The invariant mass distribution is 
symmetrical with a resonant peak at $m_\text{LQ} = 750\,\text{GeV}$. The LO result is given in figure \ref{fig:vector_tree_graph}
and shows a resonant peak height of $3.79\times10^{-2}\,\text{pb/GeV}$ and comparing this with the NLO result
for the core process, (\ref{core2}), shown in figure \ref{fig:vector_qcd_graph}, the peak height has increased 
to $5.23\times10^{-2}\,\text{pb/GeV}$ giving an enhancement of 38\% over the LO result. There is also an increase in the total
cross-section from $3.30\,\text{pb}$ to $4.49\,\text{pb}$ giving an enhancement of $36\%$.
The K-factor for the NLO QCD contributions shows more variation than with the scalar leptoquark result, but is reasonably constant
across the invariant mass distribution with an average value $\sim 1.5$.

The contributions from the additional partonic processes are given in figure \ref{fig:vector_add_graph} and also show a large enhancement to the resonant peak. 
These corrections have a resonant peak height of $5.39\times10^{-2}\,\text{pb/GeV}$ giving an enhancement of 42\% over the LO result.
There is also a large enhancement to the cross-section which is increased by $44\%$ to $4.74\,\text{pb}$.
The K-factor for the additional partonic processes is reasonably constant across the invariant mass distribution with an average
value $\sim1.5$, but shows an enhancement to the K-factor for low values of the invariant mass distribution.

Combining these corrections, the total NLO result is given in figure \ref{fig:vector_total_graph} and shows a peak height of 
$6.83\times10^{-2}\,\text{pb/GeV}$ giving a large enhancement of 80\% over the LO result. The cross-section for the total NLO 
process is increased by $80\%$ to $5.93\,\text{pb}$. 
The K-factor for the total NLO result is also reasonably constant across the invariant mass distribution with an average value 
$\sim 2$, but does show an increase to the K-factor for low values of the invariant mass distribution, which arises from the contributions from the
additional partonic processes.
These results are summarised in tables \ref{table:vector_nwa_peak} and
\ref{table:vector_nwa_cs}.

\begin{table}[!ht]
 \begin{center}
 \begin{tabular}{|l|r|r|}
  \hline
  Correction Type & Peak Height (pb/GeV) & Percentage change on LO \\ \hline
  LO & $3.79\times10^{-2}$ & - \\ \hline
  NLO QCD & $5.23\times10^{-2}$ & +38\% \\ \hline
  NLO Additional & $5.39\times10^{-2}$ & +42\% \\ \hline\hline
  NLO Total & $6.83\times10^{-2}$ & +80\% \\
  \hline
 \end{tabular}
 \caption{Summary of the CTEQ6M vector results for the NWA.}
 \label{table:vector_nwa_peak}
 \end{center}
\end{table}
\begin{table}[!ht]
 \begin{center}
 \begin{tabular}{|l|r|r|}
  \hline
  Correction Type & Total cross-section (pb) & Percentage change on LO \\ \hline
  LO & 3.30 & - \\ \hline
  NLO QCD & 4.49 & +36\% \\ \hline
  NLO Additional & 4.74 & +44\% \\ \hline\hline
  NLO Total & 5.93 & +80\% \\
  \hline
 \end{tabular}
 \caption{Summary of the CTEQ6M vector results for the NWA.}
 \label{table:vector_nwa_cs}
 \end{center}
\end{table}

\subsection{Non-Factorisable Results}
For the non-factorisable results the LO contribution is given in figure \ref{fig:vector_tree_graph} and shows the same resonant 
peak at $m_\text{LQ}=750\,\text{GeV}$ as in the NWA, but compared with the scalar leptoquark results there is now a noticeable difference 
between the peak heights for the NWA and full process. 
This is due to the fact that for a vector leptoquark there is interference between amplitudes in which the
intermediate leptoquark has different helicities.
As in the scalar case the distribution for the full process is not a symmetric and there is an enhancement to the
invariant anti-lepton jet invariant mass $m_\text{inv}$ for $m_\text{inv}<m_\text{LQ}$. There is also
an increase in the total-cross section, at LO the cross-section for the full non-factorisable process is $5.57\,\text{pb}$. This
increase is primarily due to the enhancement in the region $m_\text{inv}<m_\text{LQ}$.

The NLO QCD corrections to the core process, (\ref{core2}), are given in figure \ref{fig:vector_qcd_graph} and show a much bigger enhancement over the LO result than 
occurs in the NWA. For the full process the peak height is $6.47\times10^{-2}$ giving a large enhancement of 57\% over the LO 
result. The total cross-section also has an increase of $62\%$ to $9.03\,\text{pb}$.
The K-factor for the NLO QCD corrections shows some variation, with an increase in the K-factor for low values of the invariant
mass distribution, but is otherwise reasonably constant with an average value $\sim 1.5$.

The contributions for the additional partonic processes at NLO, shown in figure \ref{fig:vector_add_graph}, also show a much bigger enhancement over the LO result than
occurs in the NWA. For the full process the peak height is $6.56\times10^{-2}\,\text{pb/GeV}$ giving a large enhancement of 60\% 
over the LO result. The total cross-section also has an enhancement of $121\%$ to $12.29\,\text{pb}$.
The K-factor for the additional partonic processes is also reasonably constant across the invariant mass distribution and 
shows less of an increase for low values of the invariant mass than is apparent in the NWA results.

Combining all of the contributions the total NLO results are shown in figure \ref{fig:vector_total_graph} and has a resonant peak 
with a height of $8.92\times10^{-2}\,\text{pb/GeV}$ giving an enhancement of 117\% over the LO result. The cross-section for the 
total NLO contributions is increased by $183\%$ to $15.75\,\text{pb}$. 
The K-factor for the total NLO contribution shows an increase for lower values of the invariant mass distribution otherwise is
reasonably constant with an average value $\sim 2.0$.
A summary of these results is given in tables 
\ref{table:vector_full_peak} and \ref{table:vector_full_cs}.

As with the scalar leptoquarks a comparison between the NWA and non-factorisable results for the vector leptoquark can be made 
by looking at the differences between the two sets of results, see figure \ref{fig:vector_diff_graph}. In common with the scalar
results there is a difference between the NWA and non-factorisable results in the region $m_\text{inv}<m_\text{LQ}$ coming
from the enhancement to the distribution due to the non-factorisable process. The NLO result again shows the biggest difference 
which is primarily due to the additional NLO contributions. Compared with the scalar results there are also differences between
the vector NWA and non-factorisable results around the resonant peak and in the case of the NLO results this is particularly large.

\begin{table}[!ht]
 \begin{center}
 \begin{tabular}{|l|r|r|}
  \hline
  Correction Type & Peak Height (pb/GeV) & Percentage change on LO \\ \hline
  LO & $4.11\times10^{-2}$ & - \\ \hline
  NLO QCD & $6.47\times10^{-2}$ & +57\% \\ \hline
  NLO Additional & $6.56\times10^{-2}$ & +60\% \\ \hline\hline
  NLO Total & $8.92\times10^{-2}$ & +117\% \\
  \hline
 \end{tabular}
 \caption{Summary of the CTEQ6M vector results for the non-factorisable process.}
 \label{table:vector_full_peak}
 \end{center}
\end{table}
\begin{table}[!ht]
 \begin{center}
 \begin{tabular}{|l|r|r|}
  \hline
  Correction Type & Total cross-section (pb) & Percentage change on LO \\ \hline
  LO & 5.57 & - \\ \hline
  NLO QCD & 9.03 & +62\% \\ \hline
  NLO Additional & 12.29 & +121\% \\ \hline\hline
  NLO Total & 15.75 & +183\% \\
  \hline
 \end{tabular}
 \caption{Summary of the CTEQ6M vector results for the non-factorisable process.}
 \label{table:vector_full_cs}
 \end{center}
\end{table}

\subsection{Renormalization/Factorization Scale Dependence for Vector Leptoquarks}
\begin{figure}[!ht]  
 \begin{center}
  \includegraphics[scale=0.55]{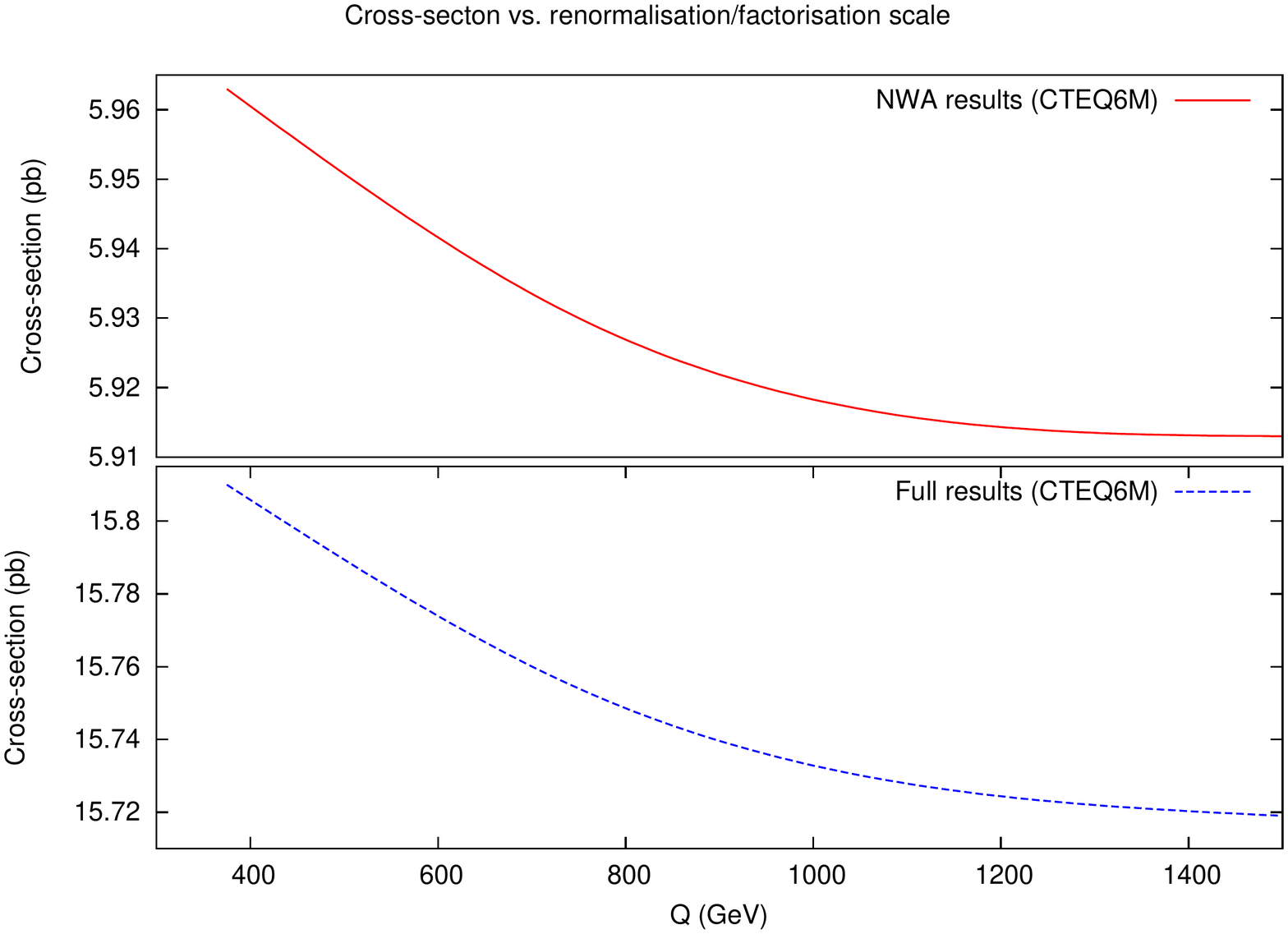}
  \caption{Scale dependence for the NWA and full non-factorisable scalar leptoquark results - with the scale $Q=\mu=\mu_F$.}
  \label{fig:vector_scale_graph}
 \end{center}
\end{figure}

As with the scalar results the scale dependence for the vector leptoquarks is also
checked. The results of the scale dependence for the NWA and full non-factorisable process are shown in figure \ref{fig:vector_scale_graph}. 
As in the case with the scalar leptoquarks the NLO results for the vector leptoquarks are also reasonably insensitive to changes in $\mu$ and $\mu_F$.

\section{Conclusions}
Having calculated the results for the NWA and the full non-factorisable process for both a scalar ($R_2$) and vector ($U_1$) leptoquark
we will conclude with a discussion and comparison of these results.

\subsection{Scalar Leptoquarks}
In general the NWA involving an intermediate scalar particle should give a good approximation to the full result close to the 
resonant region. The reason for this is there is no sum over helicities to consider and so the only approximation needed is that 
the decay-width is narrow and the intermediate propagator can be treated as a Dirac $\delta$-function. 

Comparing the scalar leptoquark results between the NWA and the full non-factorisable process (see tables \ref{table:scalar_nwa_peak} and 
\ref{table:scalar_full_peak}) the effectiveness of the NWA approximating the full non-factorisable process at NLO is confirmed. 
In particular, we see that both sets of results give good agreement between the heights of the resonant peaks - the dominant feature 
in both sets of distributions.

Away from the resonant peak there are differences in the invariant mass distributions between the NWA and full non-factorisable
process. The NWA gives a symmetric distribution around the resonant peak whereas the non-factorisable process gives a non-symmetric
distribution. In particular there is an enhancement to the distribution for values of the invariant mass 
$m_\text{inv}<m_\text{LQ}$ coming from the non-factorisable contributions to the full process. The enhancement to the distribution
away from the resonant peak does make a large contribution to the total cross-section (see tables \ref{table:scalar_nwa_cs} and 
\ref{table:scalar_full_cs}) and we see a large difference between the cross-sections calculated in the NWA and full non-factorisable
process.

\subsection{Vector Leptoquarks}
Comparing the vector leptoquark results the NWA does not provide as good an approximation to the full non-factorisable process for vector 
leptoquarks as it does for scalars. The key reason for this is because in the NWA the interference between the different helicity states of
the intermediate leptoquark are assumed to be negligible and are ignored. Looking at the difference between the resonant peak heights in
tables \ref{table:vector_nwa_peak} and \ref{table:vector_full_peak} the agreement between the two sets of results is not as close 
as in the scalar case and suggests that the helicity interference terms do make a noticeable contribution.

The anti-lepton/jet invariant mass distribution also shows the same features as in the scalar leptoquark case. The NWA gives a symmetric
distribution around the resonance, but the full non-factorisable process also shows an enhancement to the distribution for values
of the invariant mass $m_\text{inv}<m_\text{LQ}$. This enhancement makes a sizeable difference to the total cross-section
(see tables \ref{table:vector_nwa_cs} and \ref{table:vector_full_cs}). The total cross-section for the non-factorisable process is 
significantly larger than in the NWA and this increase is primarily caused by the additional NLO corrections.

In general when looking at both the scalar and vector leptoquark results there are two important features they have in common:
\begin{itemize}
 \item The NLO corrections are large compared to the LO results, particularly with regards to the full non-factorisable process, but
this is often the case when including QCD corrections.
 \item From the non-factorisable results there are substantial corrections below the resonance, particularly with regards to 
the vector leptoquarks, and its is possible these could be observed to give an indirect hint of the presence of leptoquarks.
\end{itemize}
\bigskip

\section*{Acknowledgements} The authors are grateful to Sasha Belyaev,
 Mike Seymour and Stefano Moretti for useful conversations and advice.
 One of us (JBH) wishes to thank STFC for financial support.

 \pagebreak

\end{document}